\documentclass{article}

\pdfoutput=1

\usepackage{arxiv}

\usepackage[utf8]{inputenc} 
\usepackage[T1]{fontenc}    
\usepackage{hyperref}       
\usepackage{url}            
\usepackage{booktabs}       
\usepackage{amsfonts}       
\usepackage{nicefrac}       
\usepackage{microtype}      
\usepackage{lipsum}		
\usepackage{graphicx}
\usepackage{natbib}
\usepackage{doi}

\usepackage{amsmath}
\usepackage{array}
\title{Optimal predictive probability designs for randomized biomarker-guided oncology trials}

\author{ \href{https://orcid.org/0000-0002-1402-4498}{\includegraphics[scale=0.06]{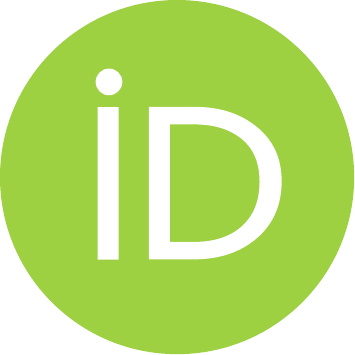}\hspace{1mm}Emily C.~Zabor} \\
	Department of Quantitative Health Sciences\\
    Taussig Cancer Institute\\
	Cleveland Clinic\\
	Cleveland, Ohio, USA \\
	\texttt{zabore2@ccf.org} \\
	\And
	\href{https://orcid.org/0000-0003-2334-5514}{\includegraphics[scale=0.06]{orcid.pdf}\hspace{1mm}Alexander M.~Kaizer} \\
	Department of Biostatistics and Informatics\\
	Colorado School of Public Health\\
    University of Colorado-Anschutz Medical Campus\\
	Aurora, Colorado, USA \\
	\texttt{ALEX.KAIZER@CUANSCHUTZ.EDU} \\
 \And
	\href{https://orcid.org/0000-0002-1458-0064}{\includegraphics[scale=0.06]{orcid.pdf}\hspace{1mm}Nathan A.~Pennell} \\
	Department of Hematology and Medical Oncology\\
	Taussig Cancer Institute\\
    Cleveland Clinic\\
	Cleveland, Ohio, USA \\
	\texttt{penneln@ccf.org} \\
 \And
	\href{https://orcid.org/0000-0003-2189-5846}{\includegraphics[scale=0.06]{orcid.pdf}\hspace{1mm}Brian P.~Hobbs} \\
	Department of Population Health\\ 
    University of Texas-Austin\\ 
    Austin, Texas, USA \\
	\texttt{brian.hobbs@austin.utexas.edu} \\
}

\renewcommand{\shorttitle}{Randomized biomarker-guided designs}

\hypersetup{
pdftitle={Randomized biomarker-guided designs},
pdfsubject={stat.AP, stat.ME},
pdfauthor={Emily C.~Zabor, Alexander M.~Kaizer, Nathan A.~Pennell, Brian P.~Hobbs},
pdfkeywords={predictive probability, oncology, clinical trial, phase II, randomized, futility monitoring, two-sample, Bayesian},
}

\begin{document}
\maketitle

\begin{abstract}
Efforts to develop biomarker-targeted anti-cancer therapies have progressed rapidly in recent years. Six antibodies acting on programmed death ligand 1 or programmed death 1 pathways were approved in 75 cancer indications between 2015 and 2021. With efforts to expedite regulatory reviews of promising therapies, several targeted cancer therapies have been granted accelerated approval on the basis of evidence acquired in single-arm phase II clinical trials. And yet, in the absence of randomization, patient prognosis for progression-free and overall survival may not have been studied under standard of care chemotherapies for emerging biomarker subpopulations prior to the submission of an accelerated approval application. Historical control rates used to design and evaluate emerging targeted therapies often arise as population averages, lacking specificity to the targeted genetic or immunophenotypic profile. Thus, historical trial results are inherently limited for inferring the potential ``comparative efficacy'' of novel targeted therapies. A recent phase III trial of atezolizumab in patients with locally advanced or metastatic urothelial carcinoma who had disease progression following platinum-containing chemotherapy found a 21.6\% response rate to standard of care chemotherapy in the biomarker subgbroup of interest, much higher than the historical control rate of 10\% that had been used to declare success in the preceding phase II trial. Consequently, randomization may be unavoidable in this setting. Innovations in design methodology are needed, however, to enable efficient implementation of randomized trials for agents that target biomarker subpopulations. This article proposes three randomized designs for early phase biomarker-guided oncology clinical trials. Each design utilizes the optimal efficiency predictive probability method to monitor multiple biomarker subpopulations for futility. A simulation study motivated by the results reported in the atezolizumab trial is used to evaluate the operating characteristics of the various designs. Our findings suggest that efficient statistical design can be conducted with randomization and futility stopping to effectively acquire more evidence pertaining to comparative efficacy before deciding to conduct a phase III confirmatory trial.
\end{abstract}

\keywords{predictive probability \and oncology \and clinical trial \and phase II \and randomized \and futility monitoring \and two-sample \and Bayesian}

\section{Introduction}

The traditional approach to oncology drug development was centered on the use of cytotoxic treatments. New treatments were evaluated in phase I dose-escalation trials to assess safety and identify the maximum tolerated dose. Next, the maximum tolerated dose would be tested for preliminary efficacy in single-arm phase II trials, with a historical control rate forming the basis of comparison. Finally, successful drugs would proceed to phase III, where randomized trials would be used to directly compare efficacy against a standard of care treatment. But drug discovery in oncology is rapidly shifting away from a focus on traditional cytotoxic treatments and toward biomarker-targeted agents. For these types of drugs, such as small molecule inhibitors, antibody drug conjugates, immune checkpoint inhibitors, and monoclonal antibodies, the traditional approach to clinical trial design has limitations. Historical control rates used in single-arm phase II studies may not be valid in the context of biomarker-targeted agents. Historical control rates used to design and evaluate emerging targeted therapies often arise as population averages, lacking specificity to the targeted genetic or immunophenotypic profile of interest. Patient prognosis for objective response, progression-free survival, and overall survival may not have been studied under standard of care chemotherapies for emerging biomarker subpopulations prior to phase III. Other factors, such as patient population drift or stage shift, add heterogeneity and bias \citep{Cannistra2009}. Consequently, expectations for response and survival for the current biomarker delineated patient populations may differ meaningfully from population averages observed in prior studies of current standard of care therapies. Additionally, in the specific context of biomarker-targeted agents, heterogeneity of response to standard of care treatments based on the biomarker of interest is also possible, so that the historical control rate may represent an averaging of effect across levels of the biomarker of interest. If the biomarker of interest is prognostic, then response to standard of care treatment in the biomarker-targeted subgroup will differ from the population-averaged response regardless of the treatment being given \citep{Ballman2015}.

Consider the recent clinical trial of atezolizumab for use in metastatic urothelial carcinoma (NCT01375842). Atezolizumab is a programmed death-ligand 1 (PD-L1) blocking monoclonal antibody that was given accelerated approval by the U.S. Food and Drug Administration in May 2016 for the treatment of patients with locally advanced or metastatic urothelial carcinoma who had disease progression following platinum-containing chemotherapy. The approval was based on the results of a single-arm phase II study in 310 patients \citep{Rosenberg2016}. The phase II study used a hierarchical fixed-sequence testing procedure to test increasingly broad subgroups of patients based on PD-L1 status, and found overall response rates of 26\% (95\% CI: 18-36), 18\% (95\% CI: 13-24), and 15\% (95\% CI 11-19) in patients with $\geq5\%$ PD-L1-positive immune cells (IC2/3 subgroup), in patients with $\geq1\%$ PD-L1-positive immune cells (IC1/2/3 subgroup), and in all patients, respectively \citep{Rosenberg2016}. All three rates exceeded the historical control rate of 10\%. Then, in March 2021, the approval in this indication was voluntarily withdrawn by the sponsor following negative results from a randomized phase III study (NCT02302807) \citep{Powles2018}. In the phase III study, 931 patients were randomly assigned to receive atezolizumab or chemotherapy in a 1:1 ratio, and the same hierarchical fixed-sequence testing procedure as in the phase II study was used. The phase III study found that overall survival did not differ significantly between the atezolizumab and chemotherapy groups of the IC2/3 subgroup (median survival 11.1 months [95\% CI: 8.6-15.5] versus 10.6 months [95\% CI: 8.4-12.2]), so no further testing was conducted for the primary endpoint \citep{Powles2018}. Further analyses revealed that while the response rates to atezolizumab were comparable to those seen in the phase II study, the response rates to chemotherapy were much higher than the historical control rate of 10\%. The overall response rates to chemotherapy were 21.6\% (95\% CI: 14.5-30.2), 14.7\% (95\% CI: 10.9-19.2), and 13.4\% (95\% CI: 10.5-16.9) for the IC2/3 subgroup, IC1/2/3 subgroup, and all patients, respectively. The overall response rates to atezolizumab were 23\% (95\% CI: 15.6-31.9), 14.1\% (95\% CI: 10.4-18.5), and 13.4\% (95\% CI: 10.5-16.9) for the IC2/3 subgroup, IC1/2/3 subgroup, and all patients, respectively. These results indicate that PD-L1 status is a prognostic biomarker, with higher response rates to both the standard of care chemotherapies that comprised the control arm and to atezolizumab treatment in the biomarker-enriched subgroup \citep{Ballman2015}. 

The example of atezolizumab in metastatic urothelial carcinoma is one of many. Between 2015 and 2021, the U.S. Food and Drug Administration approved six antibodies against PD-L1 or programmed death 1 (PD-1) for 75 cancer indications, and 35 of these approvals were accelerated based on early phase trial results \citep{Beaver2021}. This extremely rapid pace of development within a single drug class was unprecedented, and led to ten such ``dangling'' accelerated approvals, which are approved indications for which the confirmatory trial showed no benefit, yet the drug remained on the market for that indication \citep{Beaver2021}. Other voluntary withdrawals following confirmatory trial results include durvalumab treatment for metastatic urothelial carcinoma, and nivolumab and pembrolizumab treatments for metastatic small-cell lung cancer \citep{Beaver2021, Powles2020, Owonikoko2021, Spigel2021, Rudin2020}. These failed confirmatory phase III trials highlight both the need for rapid development of new treatments in patient populations with few therapeutic options, and the need for innovations that facilitate more rigorous designs of phase II trials for targeted therapies. To overcome many issues, including those associated with the use of historical control rates, randomization may be unavoidable in this setting. Arguments for the use of randomization in the phase II setting have been prominent for over a decade \citep{Korn2001, Rubinstein2005, Ratain2009, Gan2010}. In addition to addressing the inconvenient reality that historical control rates often used in single arm studies may have limited value for novel targets, randomized phase II trials can also overcome issues of selection bias and patient heterogeneity. Randomized designs that incorporate futility stopping can provide information on current control rates to the treatment under study while also stopping inefficacious treatments early. 

This article proposes three different biomarker-guided randomized phase II trial designs with optimal efficiency predictive probability monitoring for futility. Using the trial of atezolizumab for metastatic urothelial carcinoma as a case study and motivating example, we compare designs based on their traditional statistical properties of type I error and power through simulation study. The designs are also evaluated based on the number of patients enrolled, the number of patients treated, the number of patients who undergo biomarker testing, and accurate estimation of the response rates of interest. Our findings suggest that potentially smaller phase II trials to those used in practice can be designed using randomization and futility stopping to efficiently obtain more information about both the treatment and control groups prior to phase III study.

\section{Materials and Methods}

This paper focuses on the setting of a two-sample randomized trial with a binary outcome. We will refer to the binary outcome as ``response'' and use ``response rate'' to describe the probability of a response throughout the article, in line with the motivating example of the phase II study of atezolizumab in metastatic urothelial carcinoma, which estimated response rates among biomarker subpopulations and compared to the historical average in the primary analysis. Any hypothetical measure of efficacy, however, such as progression-free survival, could be used with the design methodology proposed. Each patient enrolled in the trial is denoted by $i$, and they either have a response such that $x_i = 1$ or do not have a response such that $x_i = 0$. Then $X = \sum_{i=1}^n x_i$ represents the total number of responses out of $n$ currently observed patients, up to a maximum sample size of $N$ total patients. The probability of response is denoted $p$, where $p_0$ represents the null response rate under the standard of care treatment and $p_1$ represents the alternative response rate under the experimental treatment. We wish to test the null hypothesis $H_0: p_1 \leq p_0$ versus the alternative hypothesis $H_1: p_1 > p_0$. 

The Bayesian statistical paradigm is based on a mathematical approach to combine prior distributions, which reflect prior beliefs about parameters such as the true response rate, with observed data (e.g., the observed number of responses in a given trial) to obtain posterior distributions of the model parameters. Here the prior distribution of the response rate has a beta distribution $Beta(a_0, b_0)$. We specifically use a $Beta(0.5, 0.5)$ prior distribution, which reflects the effective information of a single patient's observation. Our data $X$ follow a binomial distribution $bin(n, p)$. We combine the likelihood function for the observed data $L_x(p) \propto p^x (1-p)^{n-x}$ with the prior to obtain the posterior distribution of the response rate, which follows the beta distribution $p|x \sim Beta(a_0 + x, b_0 + n - x)$. Posterior probabilities represent the probability that the experimental response rate exceeds the null response rate based on the data accrued so far in the trial. Posterior decision can be obtained by applying a clinically relevant threshold, $\theta$, to the posterior distribution. We would declare a treatment efficacious if the posterior probability exceeded the posterior threshold, i.e. $\Pr(p_1>p_0 | X) > \theta$.

Bayesian predictive probability monitoring has been a popular approach for designing clinical trials with sequential futility monitoring \citep{Lee2008, Dmitrienko2006, Saville2014, Hobbs2018}. It is a natural fit for this type of trial, as it allows for flexibility in both the timing and the number of looks. In addition, predictive probability is an intuitive interim monitoring strategy because it tells the investigator what the chances are of declaring the treatment efficacious at the end of the trial if enrollment is continued to the maximum planned sample size. At any given interim look, the posterior predictive distribution of the number of future responses $X^*$ in the remaining $n^*=N-n$ future patients follows a beta-binomial distribution $Beta\text{-}binomial(n^*, a_0 + x, b_0 + n - x)$. The posterior predictive probability (PPP) represents the probability that the experimental treatment will be declared efficacious at the end of the trial when full enrollment is reached, conditional on the currently observed data and the specified priors. The posterior predictive probability is calculated as $PPP = \sum_{{x^*}=0}^{n^*} \Pr(X^*=x^*|x) \times I(\Pr(p_1>p_0 | X, X^*=x^*) > \theta)$. A second predictive threshold $\theta^*$ is defined, and we would stop the trial early for futility if the predictive probability dropped below the given threshold, i.e. $PPP<\theta^*$. Predictive thresholds closer to 0 lead to less frequent stopping for futility whereas predictive thresholds closer to 1 lead to frequent stopping in the absence of almost certain probability of success. 

When designing a trial with sequential predictive probability monitoring for futility, it is essential to ensure the trial conforms to traditional standards for type I error control and power. To do so, we must examine the operating characteristics of a variety of designs based on combinations of the posterior threshold $\theta$ and the predictive threshold $\theta^*$ and select a single design for use in the trial. In earlier work, we proposed two optimization criteria to help select from among a variety of designs in the setting of a one-sample study \citep{Zabor2022}. Here we focus on the optimal efficiency criteria and extend the approach to the setting of a two-sample study for targeted therapy. 

The simulation study is based on the phase II trial of atezolizumab in metastatic urothelial carcinoma. There are three independent biomarker subgroups based on the percentage of PD-L1-expressing immune cells: IC0 ($<1\%$), IC1 ($\geq1\%$ and $<5\%$), and IC2/3 ($\geq5\%$). The subgroups have equal prevalence of $33\%$ in the study population. We consider a standard of care arm denoted ``chemotherapy'' and an experimental treatment arm denoted ``atezolizumab''. The null response rate was based on the stated historical control rate of 10\% \citep{Rosenberg2016}. As no specific alternative was specified, we examine subtype-specific alternative rates of 10\%, 20\%, and 30\% in the IC0, IC1, and IC2/3 subgroups, respectively, in line with what we expect for a predictive biomarker, for which the treatment effect differs according to levels of the biomarker of interest \citep{Ballman2015}. Interim looks for futility are planned after every 10 patients. A random number of responses was generated for every 10 patients up to the maximum sample size, based on a binomial distribution with the setting-specific response rate. 1000 simulated datasets were generated under the null and 1000 simulated datasets were generated under the alternative. We considered posterior thresholds $\theta$ of 0.7, 0.74, 0.78, 0.82, 0.86, 0.9, 0.92, 0.93, 0.94, 0.95, 0.96, 0.97, 0.98, and 0.99, and predictive thresholds $\theta^*$ of 0.05, 0.1, 0.15, and 0.2. For each combination of posterior and predictive threshold, the predictive probability that the experimental treatment arm response rate exceeds the standard of care arm response rate at the end of the trial is calculated at each interim look until it either fell below the given predictive threshold or the end of the trial was reached, whichever came first. If the end of the trial was reached, the trial was considered positive if the predictive probability was greater than the given posterior threshold and negative otherwise. If halted early for futility, the trial was considered negative. We propose and compare three strategies for conducting randomized two-sample biomarker-guided designs that use optimal efficiency predictive probability monitoring for futility: a pooled control arm design, a stratified control arm design, and an enrichment design. 

\begin{figure}
\begin{center}
\includegraphics[width=0.8\textwidth]{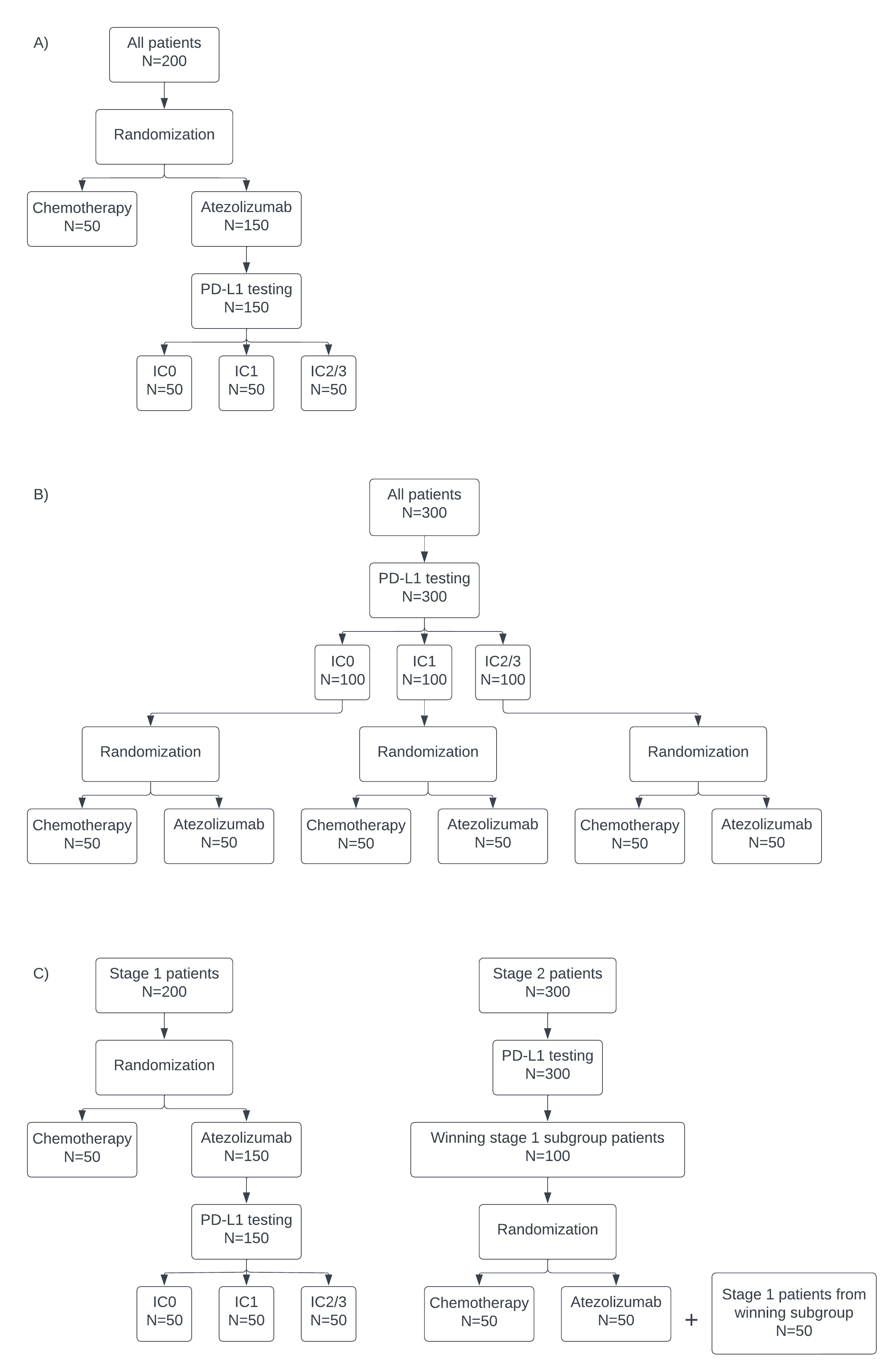}
\end{center}
\caption{Diagrams of the A) pooled randomization, B) stratified randomization, and C) enrichment designs.}
\label{fig:diagrams}
\end{figure}

The pooled control arm design is depicted in Figure~\ref{fig:diagrams}A. In this design, patients are randomized to atezolizumab or chemotherapy in a 3:1 ratio. PD-L1 testing is performed only on patients randomized to receive atezolizumab. The atezolizumab arm is separated into three biomarker-specific treatment subgroups. This design has a maximum sample size of 200: n=50 patients for the pooled chemotherapy control arm and n=50 for each PD-L1 biomarker-specific atezolizumab arm. 

The stratified control arm design is depicted in Figure~\ref{fig:diagrams}B. In this design, PD-L1 testing is conducted on all patients. Then, within each subgroup, patients are randomized to atezolizumab or chemotherapy in a 1:1 ratio. This design has a maximum sample size of 300: n=50 for each PD-L1 biomarker-specific chemotherapy and atezolizumab arm. 

The enrichment design is depicted in Figure~\ref{fig:diagrams}C. This design is equivalent to the pooled design at stage 1. If all subgroups stop for futility in stage 1, then the trial is stopped. Otherwise, at the end of stage 1, the subgroup with the highest posterior predictive probability, subject to some lower bound, continues to stage 2. The lower bound was selected as the 80th percentile of maximum posterior predictive probability across the three subgroups at stage 1 under the null. This percentile was used to target a 20\% rate of moving a subgroup forward when all of the subgroups are truly null. This higher rate of stage 1 type I error is consistent with the phase objective, which emphasizes acquiring more data on safety and efficacy for promising treatments in an early phase trial designs of this type. The actual type I error at stage 1 was calculated as the proportion of simulated trials under the global null, i.e. if all three biomarker-specific subgroups had a true response rate of 10\%, in which the subgroup with maximum posterior predictive probability exceeded the lower bound and did not stop early for futility, so was selected to continue to stage 2. The power at stage 1 was calculated as the proportion of simulated trials under the alternative in which the IC2/3 subgroup was selected as having the maximum posterior predictive probability, subject to the lower bound, and did not stop early for futility. In stage 2, PD-L1 testing is conducted on all patients. Only those patients belonging to the subgroup selected in stage 1 are enrolled on the trial and randomized 1:1 to atezolizumab or chemotherapy. The stage 1 treatment group results for the selected subgroup, if any, are carried forward into stage 2. An additional n=100 biomarker-specific patients are enrolled at stage 2, for a total maximum sample size of 300.

For the pooled and stratified designs, the type I error was calibrated in the IC0 subgroup null setting as the proportion of simulated trials in which the IC0 subgroup was declared positive as compared to the control group. The power was calibrated in the IC2/3 subgroup alternative setting as the proportion of simulated trials in which the IC2/3 subgroup was declared positive as compared to the control group. The IC1 subgroup was considered an intermediate setting and no results were calibrated based on this subgroup. For the enrichment design, the type I error was calibrated based on the stage 2 results as the proportion of simulated trials under the null in which the selected subgroup, if any, was declared positive. The power was calibrated based on the stage 2 results as the proportion of simulated trials under the alternative in which the selected subgroup, if any, was declared positive. The stage 2 calibration occurs in fewer than 1000 simulated trials, as only the specific trials in which a subgroup was selected to continue to stage 2 were used. The resulting design options were limited to those that resulted in a type I error rate between 0.05 and 0.1 and a power of at least 0.8. Then, the efficiency distance metric was calculated as described in \citet{Zabor2022} using total trial-level sample sizes. The design with the minimal efficiency distance metric was identified as the optimal design.

All results were generated using R software version 4.2.0 \citep{RCT2022} along with the `ppseq' R package \citep{Zabor2021}.

\section{Results}

\begin{figure}
\begin{center}
\includegraphics[width=0.9\textwidth]{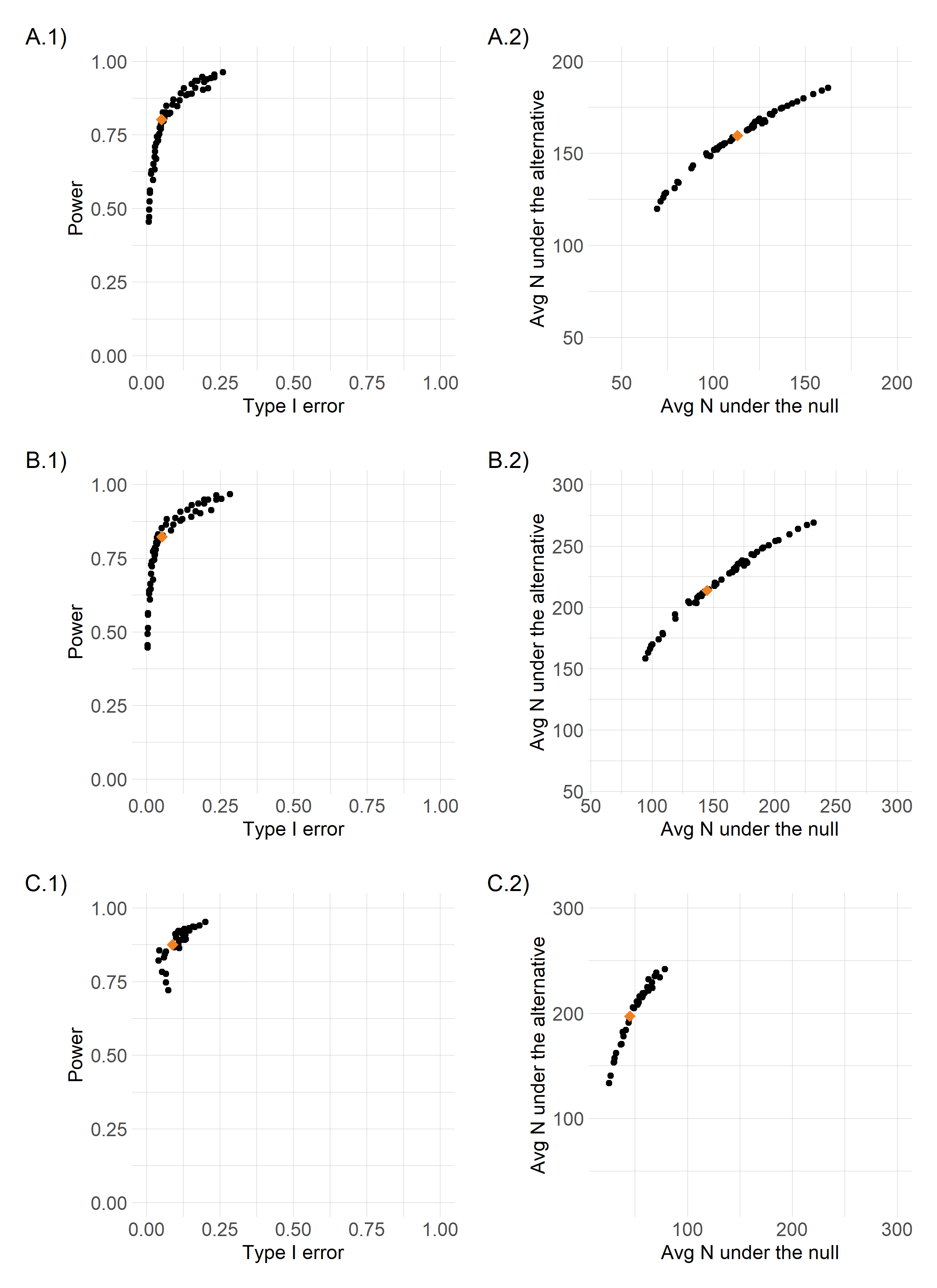}
\end{center}
\caption{Plots of design options for A) the pooled control design, B) the stratified control design, and C) the enrichment design (stage 2 results only) based on 1) accuracy defined as type I error by power and 2) efficiency defined as average total sample size under the null versus average total sample size under the alternative. The orange diamond identifies the design that was identified to have optimal efficiency.}
\label{fig:operatingchars}
\end{figure}

The accuracy and efficiency results for all 56 possible posterior and predictive threshold combinations are plotted in Figure~\ref{fig:operatingchars}. Each point represents the combination of one posterior threshold and one predictive threshold, and the orange diamond on each plot identifies the design that was identified to have optimal efficiency while maintaining type I error between 0.05 and 0.1 and with power of at least 0.8. For the enrichment design, only threshold combinations that ever proceeded to stage 2 are plotted so there are only 36 points, as 20 threshold combinations never resulted in designs that continued to stage 2. The optimal efficiency pooled control arm design had posterior threshold 0.9 and predictive threshold 0.1, the optimal efficiency stratified control arm design had posterior threshold 0.9 and predictive threshold 0.2, and the optimal efficiency enrichment design had posterior threshold 0.96 and predictive threshold 0.15. We see that the different threshold combinations result in a wide range of results, some with low power $< 0.5$ or high type I error $> 0.2$, and some with too low average sample size under the alternative or too high average sample size under the null. By applying the optimal efficiency criteria, we are able to identify a design that seeks to maximize sample size under the alternative and minimize sample size under the null, within the pre-specified range of type I error and minimum power. 

\begin{table}
\centering
\begin{tabular}{lrrrrrrrr}
  \hline
   & \multicolumn{2}{c}{Pooled} & \multicolumn{2}{c}{Stratified} &  \multicolumn{4}{c}{Enrichment} \\
   \cline{2-9}
   & \multicolumn{2}{c}{} & \multicolumn{2}{c}{} &  \multicolumn{2}{c}{Stage 1} &  \multicolumn{2}{c}{Stage 2}\\
   \cline{6-9}
Subgroup & Type I error & Power & Type I error & Power & Type I error & Power & Type I error & Power \\ 
  \hline
IC0 & 0.05 & 0.05 & 0.05 & 0.08 & -- & -- & -- & --\\ 
  IC1 & 0.05 & 0.40 & 0.08 & 0.45 & -- & -- & -- & --\\ 
  IC2/3 & 0.07 & 0.80 & 0.07 & 0.82 & -- & -- & -- & --\\ 
  Overall & -- & -- & -- & -- & 0.09 & 0.73 & 0.09 & 0.86\\
   \hline
\end{tabular}
\caption{\label{tab:acc-tab}Type 1 error and power by design and treatment subgroup}
\end{table}

The type I error and power for each biomarker-specific subgroup under the pooled control arm and stratified control arm designs, and the overall type I error and power for the enrichment design, are presented in Table~\ref{tab:acc-tab}. We see that both the pooled control arm and stratified control arm designs result in reasonable power to detect an effect for the IC2/3 subgroup, with slightly higher power of 0.82 in the stratified control arm design as compared to 0.8 in the pooled control arm design. The pooled control arm design and stratified control arm design both have type I error of 0.07 for the IC2/3. Both the pooled control arm and stratified control arm designs have very low power $< 0.5$ to detect the IC1 subgroup and $< 0.1$ to detect the IC0 subgroup. Only overall results are available for the enrichment design, which results in a type I error and power of 0.09 and 0.73, respectively, for stage 1. The type 1 error rate of 0.09 for stage 1 means that under the null 91\% of simulated trials did not proceed to stage 2; however, 4.2\% proceeded to stage 2 with the IC2/3 subgroup, 2.9\% proceeded to stage 2 with the IC1 subgroup, and 1.9\% proceeded to stage 2 with the IC0 subgroup. The power of 0.73 for stage 1 means that under the alternative 73\% of simulated trials proceeded to stage 2, and all of them did so with the IC2/3 subgroup; the remaining 27\% of simulated trials did not proceed to stage 2. The overall type I error rate for stage 2 of the enrichment design was 0.09 and the overall power for stage 2 of the enrichment design was 0.86. Since the IC2/3 subgroup was exclusively carried forward to stage 2 of the enrichment design under the alternative, this could also be considered the power for the IC2/3 subgroup, and it exceeds the power of 0.82 of the stratified control arm design and the power of 0.8 of the pooled control arm design.

\begin{table}
\centering
\begin{tabular}{lrrrrrr}
  \hline
  & \multicolumn{2}{c}{Pooled} & \multicolumn{2}{c}{Stratified} &  \multicolumn{2}{c}{Enrichment} \\
   \cline{2-7}
Subgroup & Avg N Null & Avg N Alt & Avg N Null & Avg N Alt & Avg N Null & Avg N Alt \\ 
  \hline
Control & 35.0 & 48.0 & 72.4 & 106.9 & 33.1 & 81.3 \\ 
  IC0 & 25.9 & 26.7 & 23.6 & 23.8 & -- & -- \\ 
  IC1 & 26.3 & 38.8 & 24.4 & 37.5 & -- & -- \\ 
  IC2/3 & 26.1 & 46.2 & 24.3 & 45.6 & -- & -- \\ 
  Total Atezolizumab & 78.2 & 111.7 & 72.4 & 106.9 & 131.0 & 159.0 \\
  Total Enrolled & 113.2 & 159.6 & 144.8 & 213.8 & 164.1 & 240.3 \\ 
   \hline
\end{tabular}
\caption{\label{tab:eff-tab}Average sample size under the null (``Avg N Null'') and average sample size under the alternative (``Avg N Alt'') by design and treatment subgroup}
\end{table}

The resulting average sample sizes under the null and alternative for each selected optimal efficiency design are presented in Table~\ref{tab:eff-tab}. We see that the pooled control arm design has the lowest total average sample size under the null at 113.2 as compared to 144.8 in the stratified control arm design and 164.1 in the enrichment design. However, the stratified control arm design enrolls the lowest percentage of the maximum possible total sample size under the null of $48.3\%$ of the maximum sample size of 300 as compared to $56.6\%$ of the maximum sample size of 200 for the pooled control arm design and $54.7\%$ of the maximum sample size of 300 for the enrichment design. This is because the stratified control arm design will stop both the control group and treatment group within a biomarker-specific subgroup if futility is determined, whereas the pooled control arm design and stage 1 of the enrichment design only stop the control group early if all three biomarker-specific subgroups stop early for futility. The enrichment design has the highest total average sample size under the alternative at 240.3. This reflects the low type I error at stage 1 and high power at stage 2, indicating that the design is identifying the most promising biomarker-specific subgroup yielding a high rate of stage 2 success. The enrichment design also has the highest total number of patients treated on atezolizumab under the alternative of 159.0, on average, as compared to 111.7 and 106.9, on average, in the pooled control arm and stratified control arm designs, respectively. The use of the pooled control arm at stage 1 of the enrichment design helps to maximize the number of treated patients at that stage and stage 2 frequently reaches full enrollment in the selected biomarker-specific subgroup. 

The enrichment design requires testing the largest number of patients, with 450 patients requiring PD-L1 testing if the design proceeds to stage 2, whereas the pooled design only tests 150 patients and the stratified design tests 300 patients. The pooled control arm and enrichment designs cannot address the question of whether the biomarker is predictive of response to the standard of care treatment, since they do not estimate response rates separately within each biomarker subgroup, though the enrichment design can fully characterize the response rate to standard of care treatment within the selected stage 2 biomarker subgroup. Only the stratified control arm design fully characterizes the response rates to standard of care treatment within each biomarker subgroup, and can therefore address the question of whether the biomarker is predictive of response for both standard of care and targeted therapies.

\section{Discussion}

This article presented three different optimal efficiency predictive probability designs for randomized biomarker-guided oncology clinical trials. A simulation study was conducted to demonstrate that posterior and predictive thresholds can be selected to maintain appropriate levels of type I error between 5\% and 10\% and power of at least 80\% in all three designs. This work was motivated by the case study of atezolizumab for the treatment of patients with locally advanced or metastatic urothelial carcinoma who had disease progression following platinum-containing chemotherapy. In the phase II trial that led to accelerated approval, 310 patients were enrolled and treated. The three proposed designs result in average phase II sample sizes that are 23-48\% smaller under the alternative and 47-64\% smaller under the null, and therefore represent a more efficient use of both human and financial resources. 

At the same time, the three proposed designs provide additional information about response rates to standard of care treatment in the control arms, thus potentially avoiding the pitfall of the atezolizumab trial, in which the historical control rate used to show efficacy in the phase II trial proved to be far below the actual response rate to standard of care treatment in the biomarker-targeted subgroup of patients. The stratified control arm design results in the most information, allowing one to determine if the biomarker of interest is predictive of response to either the standard of care treatment or the experimental treatment or both. The enrichment design only characterizes the response rate to the standard of care treatment in the biomarker subgroup that is selected to continue to stage 2. But both the pooled control arm design and stage 1 of the enrichment design are superior to use of a historical control rate, since the patient population of the control group is identical to that of the treatment group in both timing and characteristics as a result of randomization. So these designs not only have lower average sample sizes than the 310 used in the atezolizumab phase II trial, but also have properties such as control groups and sequential futility monitoring that facilitate valid inference of comparative effectiveness and improve decision-making for continuation to phase III. The phase III trial of atezolizumab for this patient population randomized 931 patients who could have been available to enroll in trials of more promising treatments, or could have avoided the rigors of a clinical trial altogether in favor of the established standard of care treatment.

The decision of which design to select will depend on a number of factors. One is the costs of biomarker testing, including invasiveness of the testing procedure, turnaround time, and actual financial cost. The enrichment design tests the most patients whereas the pooled control arm design tests the fewest patients. So in the case of extremely invasive or expensive tests, the pooled control arm design may be preferred. Another consideration is the prevalence of the biomarker in the population. The enrichment design in stage 2 requires testing all patients in order to identify and enroll only patients with the biomarker of interest, which could be prohibitively expensive or time consuming in the setting of a rare biomarker. In that case, the pooled control arm design may be preferable since all patients are enrolled and the control group will more easily reach full enrollment by containing a mix of patients regardless of biomarker status. But any of the proposed designs could result in a more efficient use of resources in the setting of a rare biomarker, considering both the ability to stop the trial early for futility, and the potential to avoid embarking on a confirmatory trial without adequate information about the population under study. A third consideration is clinical evidence for the biomarker being prognostic in nature, leading to differential response rates across biomarker subgroups on even standard of care therapies. If there is preliminary evidence or biological plausibility that such an effect might exist, the stratified control arm design may be preferable since it fully characterizes the response rates of the control groups within each biomarker subgroup. And a final consideration is clinical evidence for the biomarker being predictive of experimental treatment response. If there is a strong belief that only biomarker positive patients will benefit from the the treatment under study, then the enrichment design may be best as it enrolls more patients in only the selected subgroup at stage 2.

The main limitation to the use of these designs is the computational intensity required to perform calibration across a variety of posterior and predictive thresholds for the setting of interest in order to identify a design with the desired operating characteristics of type I error and power. While we have developed open-source R software for the design of single-arm and two-arm optimal sequential predictive probability designs \citep{Zabor2021}, specialized programming using the functions from the `ppseq' R package would be required to design a pooled control arm, stratified control arm, or enrichment design of the type presented here. Moreover, a large memory server is needed to complete the computations in a reasonable time span. However, once the thresholds have been selected, decision rule tables for early stopping can be generated so that no mid-trial computations would be necessary.

As rapid development of biomarker-targeted agents in oncology continues, new implementations of existing statistical methods such as those presented here will represent the most nimble way for the statistical design of trials to keep up with the changing context of cancer treatment. Randomization is an old statistical tool that has not traditionally been employed in early phase oncology clinical trials due to the sample size requirements. But in the context of increasingly large early phase clinical trials that can include hundreds of patients across multiple cancer types or multiple biomarker levels or both, randomization is no longer the constraint that it once was. This kind of efficient design also stresses the importance of mandating that all patients enrolled to biomarker-targeted trials have the biomarker of interest tested at enrollment (as opposed to only a subset of those with tissue available) so that the most accurate information about efficacy within biomarker groups can be obtained. Here we have demonstrated that it is possible to conduct randomized phase II trials with smaller sample sizes than those being used in practice for single-arm trials. Moreover, Bayesian sequential design with predictive probability yields more efficient and informative early phase clinical trial results than the standard frequentist approaches commonly implemented in practice.

\clearpage

\section*{Conflict of Interest Statement}

BH declares advising/consulting for Amgen, Bayer HealthCare Pharmaceuticals Inc., CSL Behring, Telperian, and STCube Pharmaceuticals. NP declares advising/consulting for AstraZeneca, Merck, Pfizer, Eli Lilly/LOXO, Genentech, BMS, Amgen, Mirati, Inivata, G1 Therapeutics, Viosera, Xencor, Janssen, Boehringer Ingelheim, Sanofi-Genzyme. NP declares advising/consulting for Merck, Pfizer, Eli Lilly, Genentech, Mirati, Janssen, Sanofi-Genzyme. EZ and AK declare no commercial or financial relationships that could be construed as a potential conflict of interest.

\section*{Author Contributions}

EZ, AK, and BP contributed to conception and design of the study. EZ performed the statistical analysis and wrote the first draft of the manuscript. All authors wrote sections of the manuscript, contributed to manuscript revision, and approved the submitted manuscript. 

\section*{Funding}

AK is supported by NHLBI K01HL151754.

\section*{Supplemental Data}

Supplementary code files and the associated results are available for all analyses included in this manuscript on GitHub at: \url{https://github.com/zabore/manuscript-code-repository/tree/master/Zabor_Randomized-Biomarker-Guided-Designs/analysis}. See the README file at \url{https://github.com/zabore/manuscript-code-repository} for details.

\section*{Data Availability Statement}

The simulated datasets and the R scripts that produced them can be accessed on GitHub at: \url{https://github.com/zabore/manuscript-code-repository/tree/master/Zabor_Randomized-Biomarker-Guided-Designs/data}. See the README file at \url{https://github.com/zabore/manuscript-code-repository} for details.\textbf{}

\clearpage


\bibliographystyle{unsrtnat}
\bibliography{ppseq_paper_references}  

\end{document}